\documentclass[aps,prb,citeautoscript,twocolumn,superscriptaddress,showkeys,amsmath,amssymb,nofootinbib]{revtex4-2}

\usepackage[caption=false,labelformat=parens]{subfig}
\usepackage{graphicx}
\usepackage{dcolumn}
\newcolumntype{.}{D{.}{.}{-1}}
\AtBeginDocument{\usepackage{booktabs}}
\usepackage{siunitx}
\usepackage{amsmath,bm,bbm}
\usepackage{microtype}
\usepackage{braket}

\usepackage{etoolbox}
\usepackage[modulo,columnwise]{lineno}
\usepackage{xcolor}
\definecolor{darkblue}{rgb}{0.33, 0.33, 0.33}
\definecolor{citeblue}{rgb}{0.2305, 0.4102, 0.9}
\usepackage{hyperref}
\hypersetup{colorlinks=true,citecolor=citeblue,urlcolor=citeblue,linkcolor=black}
\usepackage{memhfixc}

\newcommand{\GM}{$\overline{\Gamma \text{M}}$}
\newcommand{\GK}{$\overline{\Gamma \text{K}}$}
\newcommand{\SA}{Bi$_2$Te$_2$Se}

\begin{document} 

\title{Observation of Dirac Charge Density Waves in Bi$_2$Te$_2$Se}

\author{Adrian Ruckhofer}
\affiliation{Institute of Experimental Physics, Graz University of Technology, Graz, Austria}
\author{Giorgio Benedek}
\affiliation{Dipartimento di Scienza dei Materiali, Universit\`{a} degli Studi di Milano-Bicocca, Milano, Italy}
\affiliation{Donostia International Physics Center, University of the Basque Country, Donostia, San Sebastian, Spain}
\author{Martin Bremholm}
\affiliation{Centre for Materials Crystallography, Department of Chemistry and iNANO, Aarhus University, Aarhus, Denmark}
\author{Wolfgang E. Ernst}
\affiliation{Institute of Experimental Physics, Graz University of Technology, Graz, Austria}
\author{Anton Tamt\"{o}gl}
\email{tamtoegl@tugraz.at}
\affiliation{Institute of Experimental Physics, Graz University of Technology, Graz, Austria}

\begin{abstract}
While parallel segments in the Fermi level contours, often found at the surfaces of topological insulators (TIs) would imply "strong" nesting conditions, the existence of charge density waves (CDWs) - periodic modulations of the electron density - has not been verified up to now. Here, we report the observation of a CDW at the surface of the TI Bi$_2$Te$_2$Se(111), below $ \approx 350\,$K by helium atom scattering, and thus experimental evidence for a CDW involving Dirac topological electrons. Deviations of the order parameter observed below $180\,$K and a low temperature break of time reversal symmetry suggest the onset of a spin density wave with the same period as the CDW in presence of a prominent electron-phonon interaction originating from the Rashba spin-orbit coupling.
\end{abstract}

\maketitle

\maketitle 

\section{Introduction}
\label{sec:intro}
Charge density waves (CDWs) - periodic modulations of the electron density - are an ubiquitous phenomenon in crystalline metals and are often observed in layered or low-dimensional materials \cite{Inosov2008,Arnold2017,Hall2019,Wang2020,Shi2021,Tang2019}. CDWs are commonly described by a Peierls’ transition in a one-dimensional chain of atoms which allows for an opening of an electronic gap at the nesting wavevector causing a Fermi-surface nesting. However, it has been questioned whether the concept of nesting is essential for the understanding of the CDW formation \cite{Johannes2008,Inosov2008}. Instead, CDWs are often driven by strong electronic correlations and wavevector-dependent electron-phonon (e-ph) coupling \cite{Rossnagel2011}. Similarly, a nesting of sections of the Fermi surface can induce a periodic spin-density modulation, a spin-density wave (SDW) \cite{Gabovich2001}. The possibility of a simultaneous appearance of both CDW and SDW order has been studied theoretically in earlier works \cite{Fishman1992}, and a SDW was recently predicted for Weyl semimetals \cite{Kundu2021}.\\
The material class of topological insulators (TIs) has recently attracted extensive attention in a different context \cite{Bradlyn2017,Bansil2016,Hasan2015,Ando2013,Zhang2011,Hasan2010}, due to their unique electronic surface states which involve a Dirac cone with spin-momentum locking \cite{Moore2010,Chen2009}. Here we report for the first time, experimental evidence for a Dirac CDW on the surface of the TI \SA , i.e. a CDW involving Dirac topological electrons. Atom scattering experiments reveal a CDW transition temperature $T_{CDW}=\SI{350}{\K}$ of the Dirac two-dimensional electron gas (2DEG). The break of time-reversal symmetry of the CDW diffraction peaks observed at low temperature suggests a prominent role of the Rashba spin-orbit coupling with the possible onset of a SDW below \SI{180}{\K}.\\
Archetypal TIs such as the bismuth chalcogenides share many similarities with common CDW materials such as a layered structure (see \autoref{fig:Fig1}(a)) \cite{Tamtogl2018a}. The hexagonal contours at the Fermi level often found in TIs also imply a strong nesting which has led to speculations about the existence of CDWs in TIs. Furthermore, the importance of charge order in the context of unconventional superconductivity in these systems has been subject to recent studies \cite{Lawson2016,Kevy2021}. On the other hand, compared to other CDW materials, the topological surface states (TSS) of TIs such as \SA\ exhibit a characteristic spin polarisation, as studied together with the helical spin texture for the present material in refs. \cite{Miyamoto2012,Nurmamat2013}. Based on helium atom scattering (HAS) it was recently shown that periodic charge density modulations of the semimetal Sb(111) derive from multivalley charge density waves (MV-CDW) due to surface pocket states\cite{Tamtogl2019}. While MV-CDWs are generally stabilised by electron-phonon (e-ph) interaction, different mechanisms can be responsible for a CDW formation\cite{Johannes2008,Inosov2008,Rossnagel2011,Zhu2015,Ghim2022}.

\subsection{Electronic structure and electron-phonon coupling}
\begin{figure}[htbp]
	\centering
	\centering
	\includegraphics[width=0.45\textwidth]{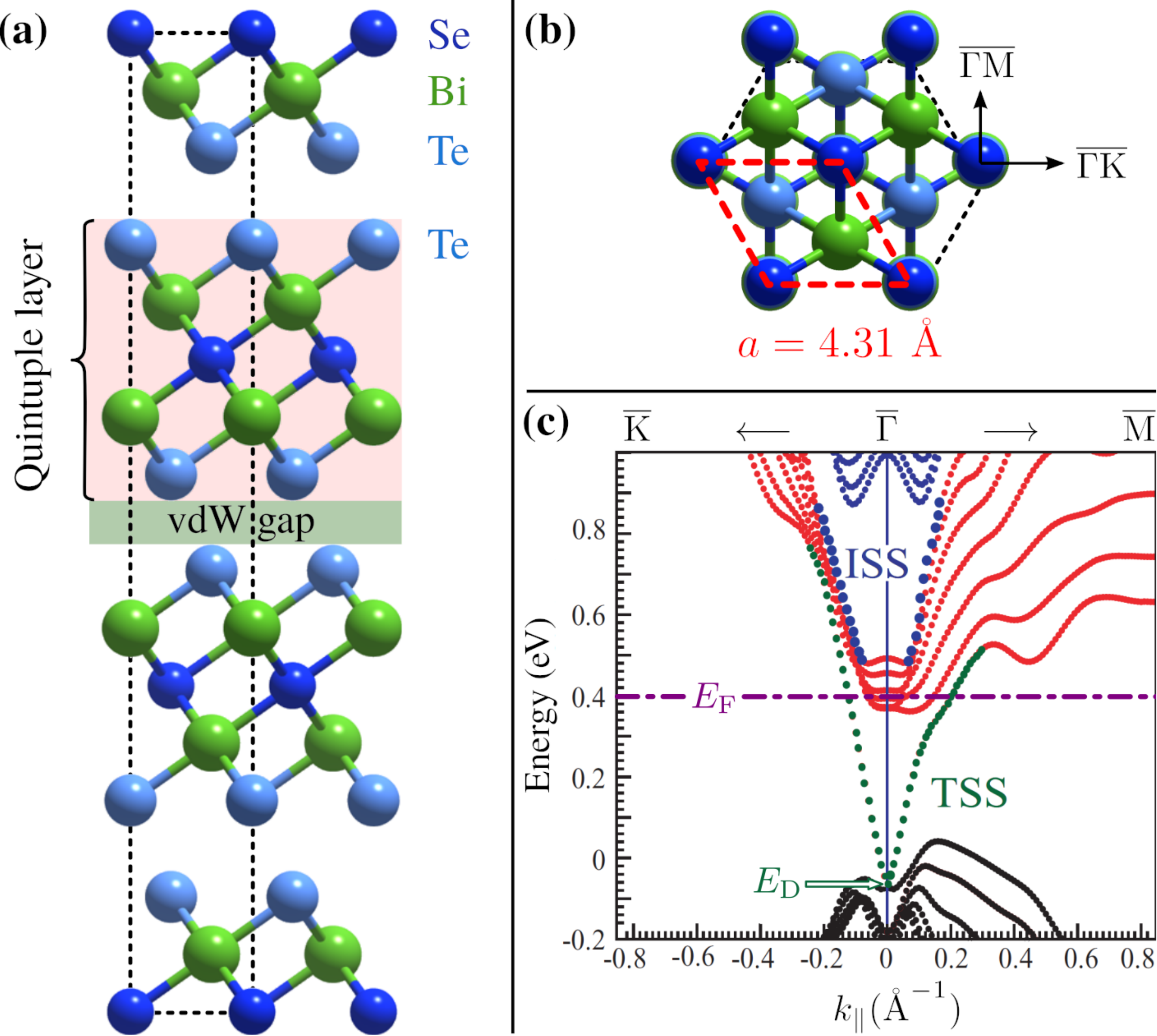}
	\caption{(a) Side view of the conventional hexagonal unit cell of \SA. The unit cell consists of three quintuple layers of which each one is terminated by a Te layer. (b) The (111) cleavage plane of \SA\ with the red rhombus highlighting the hexagonal surface unit cell. (c) Surface band structure of \SA(111) calculated by Nurmamat \textit{et al.} \cite{Nurmamat2013} (reproduced with permission, copyright 2013 by the American Physical Society) along the symmetry directions \GK\ and \GM. TSS labels the topological surface states forming the Dirac cone, while internal (quantum-well) surface states (ISS) are also found in the gap above the conduction band minimum. In the present sample the Fermi level $E_\mathrm{F}$ (purple dash-dotted line) is located about \SI{0.4}{\eV} above the Dirac point $E_\mathrm{D}$.}
	\label{fig:Fig1}
\end{figure}

Among the bismuth chalcogenides, \SA\ is much less studied. Surface dominated electronic transport \cite{Ren2010,Geh2012,Barreto2014,Shekhar2014,Cao2014} as well as the surface electronic band structure \cite{Arakane2012,Neupane2012,Miyamoto2012,Maas2014,Frantzeskakis2015} have been subject to several investigations. Moreover, in terms of the electronic band structure it was shown that for different Bi$_{2-x}$Sb$_x$Te$_{3-y}$Se$_y$ compositions, the Dirac point ($E_\mathrm{D}$ in \autoref{fig:Fig1}(c)) moves up in energy with increasing $x$ \cite{Arakane2012}. Tuning these stoichiometric properties and the doping of materials may give rise to nesting conditions between electron hole/pocket states at the Fermi surface ($E_\mathrm{F}$ in \autoref{fig:Fig1}(c)).

\noindent \autoref{fig:Fig1}(c) depicts the electronic surface band structure calculated by Nurmamat \textit{et al.} \cite{Nurmamat2013} along the symmetry directions \GK\ and \GM\ revealing the TSS which form the Dirac cone. The Fermi level $E_F$ (purple line) in our present sample is located about \SI{0.4}{\eV} above the Dirac point yielding the formation of quantum well states at the Fermi surface. Moreover, a near-surface 2DEG with a pronounced spin-orbit splitting can be induced on \SA\ by adsorption of rubidium \cite{Michiardi2015}, and surface oxidation may occur at step edge defects after cleaving \cite{Thomas2016}, but \SA\ seems to be less prone to the formation of a 2DEG from rest gas adsorption compared to other TIs \cite{Frantzeskakis2017} as shown in angle resolved photo-emission measurements of the present samples \cite{Barreto2014}. Dirac fermion dynamics in \SA\ was subject to a recent study by Papalazarou \emph{et al.} \cite{Papalazarou2018}.\\
One reason for \SA\ being less studied than the binary bismuth chalcogenides might be the difficulty in synthesising high-quality single crystals, which originates from the internal features of the specific solid-state composition and phase separation in \SA\ \cite{Mi2013}. In this work we present a helium atom scattering (HAS) study of \SA\ (phase II of Bi$_2$Te$_{3-x}$Se$_x$(111) with $x \approx  1$\footnote{According to the surface lattice constant $a = 4.31~\mbox{\AA}$ as determined by HAS and Fig. 1(b) of \cite{Mi2013}, $x \approx  1$ for the present sample.} and  approximated as \SA\ in the following). As helium atoms are scattered off the surface electronic charge distribution, HAS\cite{Farias1998,Bracco2013} provides access to the surface electron density \cite{Tamtogl2019,Holst2021a}, and is therefore a perfect probe for experimental studies of TIs, since the TSS properties are often mixed up with those of bulk-states \cite{Viti2016,Knispel2017}. Since the surface electronic transport properties of TIs at finite temperature as well as the appearance of CDWs are influenced by the interaction of electrons with phonons, the e-ph coupling as described in terms of the mass-enhancement factor $\lambda$ has been subject to several studies \cite{Hatch2011,Sobota2014,Heid2017,Tamtogl2017b,In2018,Benedek2020,Anemone2021}. For \SA\ it was reported that the electron-disorder interaction dominates scattering processes with $\lambda = 0.12$ \cite{Chen2013}, in good agreement with the value found from HAS \cite{Benedek2020}. 

\section{Experimental Details}
The experimental data of this work was obtained at a HAS apparatus, where a nearly monochromatic beam of $^4$He is scattered off the sample surface in a fixed source-sample-detector geometry (for further experimental details  see Ref. \cite{Tamtogl2010} and supplementary information). The scattered intensity of a He beam in the range of $10-15~\mbox{meV}$ is then recorded as a function of the incident angle $\vartheta_i$ with respect to the surface normal, which can be modified by rotating the sample in the scattering chamber. The momentum transfer parallel to the surface $\Delta K$, upon elastic scattering, is given by
\begin{equation}
\Delta K = \lvert \mathbf{k}_i \rvert  \left( \sin \vartheta_f  - \sin \vartheta_i  \right),
\label{eq:ExpDK}
\end{equation}
with $\mathbf{k}_i$, the incident wavevector, and $\vartheta_i$ and $\vartheta_f$ the incident and final angle, respectively. The \SA\ sample was cleaved \textit{in situ}, in a load-lock chamber \cite{Tamtogl2016a}, prior to the experiments. Due to the weak bonding between the quintuple layers in \autoref{fig:Fig1}(a), the latter gives access to the (111) cleavage plane with a Te termination at the surface (\autoref{fig:Fig1}(b)). 

\section{Results and Discussion}
\subsection{Surface CDW order}
\begin{figure}[htbp]
     \centering
     \centering
	\includegraphics[width=0.4\textwidth]{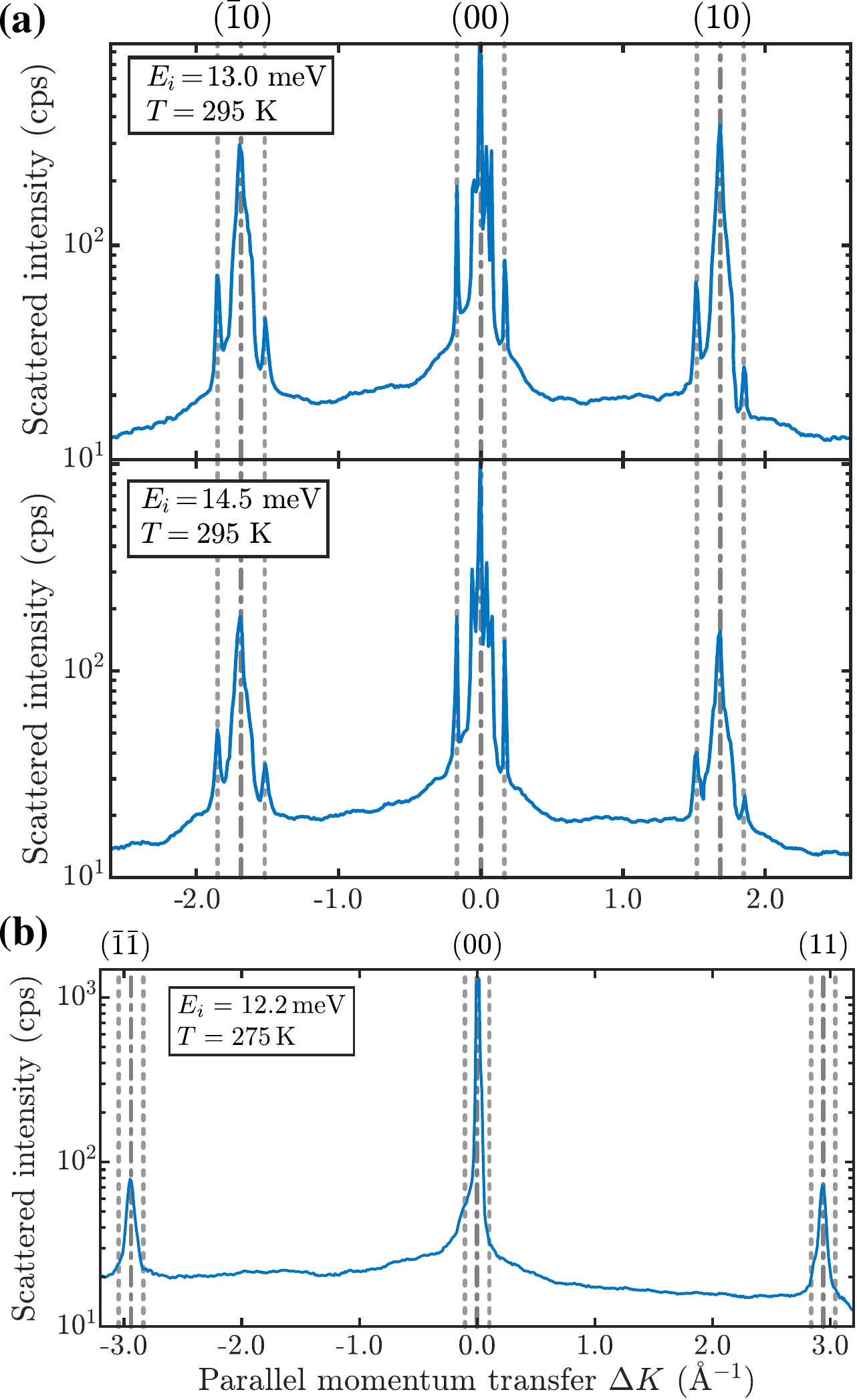}
	\caption{The CDW periodicity becomes evident in diffraction scans (logarithmic scale) of \SA(111). (a) Scans along \GM\ taken at room temperature show satellite peaks (illustrated by the grey dashed vertical lines) next to the specular and first order diffraction peaks (grey dash-dotted vertical lines). (b) Same for the \GK\ direction; in this direction no evident satellite peaks are observed close to the diffraction peaks.}
     \label{fig:Fig2}
\end{figure}

\begin{figure}[htbp]
	\centering
	\centering
	\includegraphics[width=0.41\textwidth]{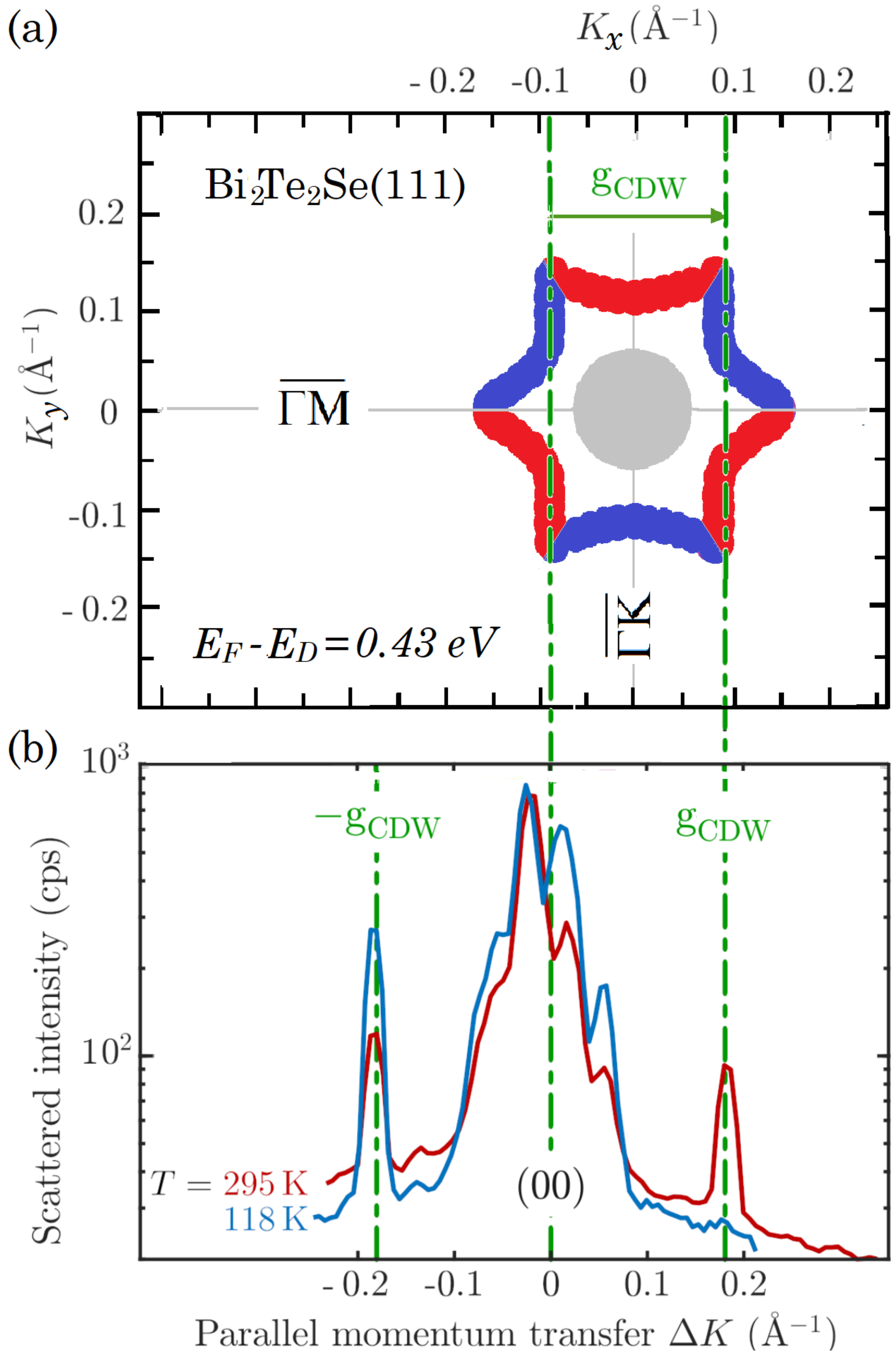}
	\caption{The spin-allowed nesting condition giving rise to the sharp CDW satellite peaks at $\pm \mathbf{g}_{\mathrm{CDW}}$ around the $(00)$-reflection is confirmed by \emph{ab initio} calculations. Due to the hexastar shape of the spin polarisation perpendicular to the surface of \SA(111) in (b)  (reproduced with permission from Nurmamat \textit{et al.} \cite{Nurmamat2013}, copyright 2013 by the American Physical Society) with red and blue circles denoting positive and negative spin polarisation, almost parallel segments with equal spin provide the nesting condition along \GM . It shows excellent agreement with the satellite structure occurring at $\pm \mathbf{g}_{\mathrm{CDW}}$ around the specular peak of \SA(111) measured with HAS in the \GM\ direction at room temperature (295 K) and plotted as red line in (c). Qualitative agreement is also found with Fourier transformed differential conductance (DC) maps, i.e. the hexagonally-symmetrised map of \SA(111) by Nurmamat \textit{et al.} \cite{Nurmamat2013} in (a).
	The nestings in the \GK\ direction, on the other hand, would imply spin reversal and no CDW satellite peaks are detected by HAS in that direction (see \autoref{fig:Fig2}(c)). The missing CDW peak at low temperature (118 K, blue line) on the right-hand side at $+\mathbf{g}_{\mathrm{CDW}}$ is attributed to the Rashba effect after the onset of a SDW spin ordering. Since the hexagonal symmetrisation of the DC map in (a) removes the effect of possible time-reversal symmetry breaking, the comparison is to be made with the room temperature HAS diffraction scan in (b). The complex structure aside the specular peak may be associated with transitions involving surface quantum-well states (see \autoref{fig:Fig1}(c) and \autoref{fig:Fig2}(a)).}
	\label{fig:Fig3}
\end{figure}

\autoref{fig:Fig2} shows the scattered He intensity versus momentum transfer $\Delta K$ \eqref{eq:ExpDK} with the scans taken at different incident beam energies $E_i$ along the \GM\ (a) and \GK\ (b) direction. Along \GM\ there appear sharp additional peaks (illustrated by the vertical dashed lines) next to both the specular and the first order diffraction peaks (vertical dash-dotted lines) at an average spacing of about $0.18~\mbox{\AA}^{-1}$ with respect to the diffraction peaks (\autoref{fig:Fig2} and \autoref{fig:Fig3}(c) in an enlarged scale). The fact that these satellite peaks appear at the same momentum transfer $\Delta K \approx \pm 0.18~\mbox{\AA}^{-1}$ with respect to the specular as well as to the first order diffraction peaks, independently of the incident energy, shows that they are neither caused by bound-state resonances\cite{Hofmann2019} nor other artifacts (see the section on CDW satellite peaks as well as additional scans in the SI), but have to be necessarily ascribed to a long-period surface superstructure of the electron density, i.e., a surface CDW.\\
For a triangular set of electron pockets with the same spin at the Fermi level, each one located at a distance $k_F$ from the zone centre, the inter-pocket nesting gives rise to satellite peaks at $\Delta K = \tfrac{3}{2} k_F$ in the \GM\ direction and $\tfrac{\sqrt{3}}{2} k_F$ in the \GK\ direction. The CDW diffraction peaks are, however, observed only in the \GM\ direction, and give $k_\mathrm{F} = 0.12~\mbox{\AA}^{-1}$. As illustrated in \autoref{fig:Fig3}, our diffraction data is best compared with calculations of the electronic structure including the spin texture. \autoref{fig:Fig3}(b) shows \emph{ab initio} calculations of the topological bands by Nurmamat \textit{et al.} \cite{Nurmamat2013} which evolve into a hexastar for a Fermi level position as shown in \autoref{fig:Fig1}(c). Due to the hexastar shape, almost parallel segments with equal spin (illustrated in blue and red) exist along \GM\ giving rise to the $\pm \mathbf{g}_{\mathrm{CDW}}$ satellite peaks in \autoref{fig:Fig3}(c).

\autoref{fig:Fig3} shows that the hexastar shape provides a spin allowed nesting which corresponds to the observed $\mathbf{g}_{\mathrm{CDW}}$ intervalley transitions. In contrast to the hexastar, for a hexagonal shape as found in several TIs or as also observed on Bi(111), opposite sides of the Fermi contour exhibit opposite spins: A situation which forbids the pairing needed for a CDW formation but leaves the possibility of a SDW\cite{Pascual2004,Kim2005}. Finally, the transmission of momentum and energy to the lattice for the spin-allowed transition across the hexastar occurs, via the excitation of virtual electron-hole pairs, if one assumes an Esbjerg-N{\o}rskov form of the atom-surface potential based on a conducting surface.\\
Qualitative agreement is also found with the Fourier transform of scanning tunnelling microscopy (STM) DC maps in reciprocal space as measured by Nurmamat \textit{et al.} \cite{Nurmamat2013} in \autoref{fig:Fig3}(a): The symmetrised DC map with a slight change of the wavevector scale (corresponding to a DC bias of $0.36\,$eV) reflects the same nesting vector as found for the room temperature data (red curve in  \autoref{fig:Fig3}(c)). The low temperature disappearance of the $+\mathbf{g}_{\mathrm{CDW}}$ peak is then explained below.

\subsection{CDW temperature dependence}
\begin{figure}[htbp]
	\centering
	\includegraphics[width=0.49\textwidth]{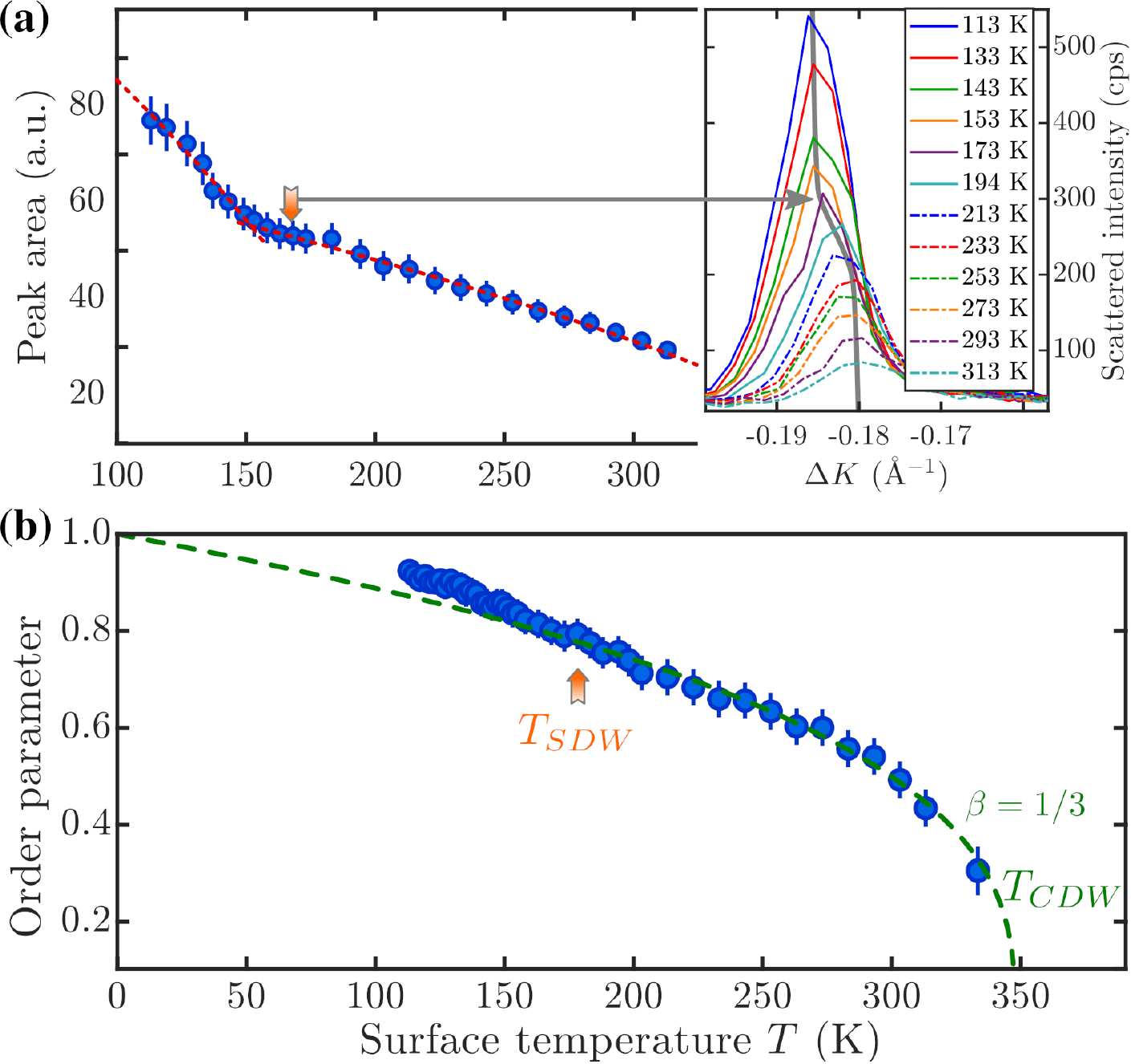}
	\caption{Peak intensity and order parameter of the surface CDW (blue circles), as extracted from the temperature dependence of the $-\mathbf{g}_{\mathrm{CDW}}$ CDW peak in \autoref{fig:Fig3}(c) (see text). The right panel in (a) shows several scans, illustrating a rapid shift of the satellite peak, occurring at around $180\,$K, from a momentum transfer of $0.18\,\mbox{\AA}^{-1}$ to $0.186\,\mbox{\AA}^{-1}$. Together with the corresponding deviation of the order-parameter fit below about $180\,$K ($T_{SDW}$) in (b) and the time-reversal symmetry break (missing $+\mathbf{g}_{\mathrm{CDW}}$ peak at low $T$ in \autoref{fig:Fig3}(c)), it suggests the onset of a spin ordering (a spin-density wave), which allows, through the Rashba effect, for a parallel momentum transfer only in one direction. (b) The fit of the order parameter with the fluctuation critical exponent $\beta = 1/3$ \cite{liu2019} (green dashed line) yields a critical temperature $T_{CDW}$ of about $350\,$K.}
	\label{fig:Fig4}
\end{figure}

In the following, we consider the temperature dependence of the CDW diffraction peaks and the CDW critical temperature $T_{CDW}$. Upon measuring the scattered intensities as a function of surface temperature it turns out that the intensity of the satellite peaks decreases much faster than the intensity of the specular peak. As shown in the supplementary information (Figure S2), when plotting both peaks in a Debye-Waller plot, the slope of the satellite peak is clearly steeper than the one of the specular peak.\\
Based on the theory of classical CDW systems, the square root of the integrated peak intensity can be viewed as the order parameter of a CDW \cite{Gruner1994,Gruner1988}. \autoref{fig:Fig4}(b) shows the temperature dependence of the square root of the integrated intensity for the $-\mathbf{g}_{\mathrm{CDW}}$ peak on the left-hand side of the specular peak (see right panel of \autoref{fig:Fig4}(b) for several scans). In order to access the intensity change relevant to the CDW system as opposed to the intensity changes due to the Debye-Waller factor, the integrated intensity $I(T)$ has been normalised to that of the specular beam \cite{Hofmann2019} - a correction which is necessary in view of the low surface Debye temperature of \SA(111)\ \cite{Benedek2020}. The temperature dependence of the order parameter $\sqrt{I(T)}$ can be used to determine the CDW transition temperature $T_{CDW}$ and the critical exponent $\beta$ belonging to the phase transition by fitting the power law 
\begin{equation}
\sqrt{ \tfrac{I(T)}{I(0)}} = \left( 1 - \tfrac{T}{T_{CDW}} \right)^\beta \; ,
\label{eq:orderpar}
\end{equation}
to the data points in \autoref{fig:Fig4}. Here, $I(0)$ is the extrapolated intensity at $0\,\mbox{K}$. The fit is represented by the green dashed line in \autoref{fig:Fig4}, resulting in $T_{CDW} = (350~\pm 10)$K and $\beta = (0.34 \pm 0.02)$.\\
The exact peak position and width of the satellite peak was determined by fitting a single Gaussian to the experimental data. The right panel of \autoref{fig:Fig4}(a) shows a shift of the satellite peak position to the right with increasing surface temperature, i.e. $\mathbf{g}_{\textrm{CDW}}$ decreases with increasing temperature, as illustrated by the grey line. Such a temperature dependence confirms the connection of the satellite peaks with the surface electronic structure. A shift of the Dirac point to lower binding energies with increasing temperature and thus a decrease of $k_F$ has been observed both for \SA(111) \cite{Nayak2018} as well as for Bi$_2$Se$_3$(111) \cite{Pan2012}. As reported by Nayak \emph{et al.} \cite{Nayak2018}, the temperature dependent changes of the electronic structure at $E_F$ occur due to the shift of the chemical potential in the case of $n$-type \SA(111). Moreover, a strong temperature dependence of the chemical potential has also been observed for other CDW systems \cite{Monney2010,Liu2022} and semiconductors \cite{Fukutani2016}.

\subsection{Diffraction and the role of spin-orbit coupling}
The surprising disappearance of the $+\mathbf{g}_{\mathrm{CDW}}$ diffraction peak observed at low temperature (118 K) is likely to be related to the apparent phase transition occurring at about 180 K (illustrated by the orange arrow in \autoref{fig:Fig4}), possibly a spin ordering within the CDW, i.e., a SDW with the same period. The latter is indicated by a rapid shift of the $-\mathbf{g}_{\mathrm{CDW}}$ diffraction peak position around 180 K, corresponding to a slight contraction of the CDW period (right panel of \autoref{fig:Fig4}(a)). The CDW order parameter, expressed by $\left(1-T/T_{CDW}\right)^\beta$ actually shows a small deviation from this law below about $180\,$K ($T_{SDW}$ in \autoref{fig:Fig4}(b)). As explained above, the He-atom diffraction process from the CDW occurs via parallel momentum transfer to the surface electron gas, via an electron-hole excitation between a filled and an empty state of equal spin and well nested at the Fermi level. Thus, the unidirectionality of the process at low temperature suggests a prominent role of the Rashba term in presence of spin ordering and strong implications for the e-ph contribution.\\
The latter is in line with Guan \textit{et al.} \cite{Guan2011}, who reported a large enhancement of the e-ph coupling in the Rashba-split state of the Bi/Ag(111) surface alloy. The larger overlap of He atoms with CDW maxima also selects electrons with the same spin, because the SDW and CDW exhibit the same period. Considering in the present case a free-electron Hamiltonian \cite{Verpoort2020},
\begin{equation}
-E_\mathrm{D} +  \dfrac{\mathbf{p}^2}{2m^* } +\alpha_R\,\bm{\sigma}\cdot \big( \mathbf{p}\times\mathbf{\hat{z}} \big)\, ,
\label{eq:hamiltonian}
\end{equation}
where $-E_\mathrm{D}$ is the energy of the Dirac point below the Fermi energy $E_\mathrm{F}=0$, $\mathbf{p}$ the surface electron momentum, and $m^*$ its effective mass, $\bm{\sigma}$ the spin operator, $\mathbf{\hat{z}}$ the unit vector normal to the surface and $\alpha_R$ the Rashba constant. The modulation of the Rashba term
\begin{equation}
\dfrac{\partial \alpha_R}{\partial A_{\mathbf{q}s}}\,A_{\mathbf{q}s}\,\bm{\sigma}\cdot \big( \mathbf{\mathbf{q}}\times\mathbf{\hat{z}} \big) \, ,
\label{eq:rashba}
\end{equation}
produced by a phonon of momentum $\mathbf{q}$, branch index $s$ and normal mode coordinate $A_{\mathbf{q}s}$ is viewed as the main source of e-ph interaction \cite{Kandemir2017}, causing interpocket coupling $(\Delta\mathbf{p}=\mathbf{q}=\pm\mathbf{g}_{\mathrm{CDW}})$ and the CDW gap opening.\\
In a diffraction process the exchange of parallel momentum between the scattered atom and the solid centre-of-mass is mediated by a virtual electron interpocket transition $\ket{\mathbf{k},n} \rightarrow \ket{\mathbf{k}+\mathbf{q},n'}$ weighed by the difference in Fermi-Dirac occupation numbers $f_{\mathbf{q},n}-f_{\mathbf{k}+\mathbf{q},n'}$. While the process $\mathbf{q} = -\mathbf{g}_{\mathrm{CDW}}$ virtually casts the electron from a pocket ground state at the Fermi level to an empty excited state across that gap, the process $\mathbf{q} = +\mathbf{g}_{\mathrm{CDW}}$ would virtually send the electron, because of the Rashba term, to a lower energy state and is therefore forbidden at low temperature.\\
Such a scenario is equivalent to saying that the SDW-CDW entanglement makes HAS sensitive to the spin orientation via its temperature dependence. The interpocket electron transition across the gap accompanying a CDW diffraction of He atoms via the modulation of the Rashba term may only occur in one direction. Since the gap energy is of the order of room temperature and at this temperature the spin ordering is removed, the above selection rule is relaxed and the diffraction peaks are observed in both directions (on both sides of the specular in \autoref{fig:Fig3}(b)).\\
The corresponding time-reversal symmetry break should actually be reflected in as-measured DC maps at low temperature, but cannot be tested based on existing STM data \cite{Nurmamat2013}, since they have been reported after hexagonal symmetrisation. Moreover, the size of the CDW gap, as inferred from the slight asymmetry between $+ \mathbf{g}_{\mathrm{CDW}}$ and $- \mathbf{g}_{\mathrm{CDW}}$, does not seem to be large enough to be resolved in ARPES\cite{Liu2022,Li2020,Fan2017}. On the other hand the nesting condition and the changes between different Fermi level contours (hexagonal vs. hexastar shape) depend strongly on any shifts of the Fermi level\cite{Frantzeskakis2017,Zunger2021,Nowak2022} and may thus be highly sensitive to the doping situation of the specific sample\cite{Fan2017}. HAS satellite diffraction peaks are much to evident, and clearly indicate an additional long-period component of the surface charge density corrugation.

\section{Summary and Conclusion}
In summary, we have provided evidence by means of helium atom scattering, of a surface charge density wave in \SA(111) occurring below 345 K and involving Dirac topological electrons. The CDW diffraction pattern is found to reflect a spin-allowed nesting across the hexastar contour at the Fermi level of previously reported \emph{ab initio} calculations\cite{Nurmamat2013}. The CDW order parameter has been measured down to $108\,$K and found to have a critical exponent of $1/3$. The observation of a time-reversal symmetry break at low temperature, together with deviations from the critical behaviour below about $180\,$K, are interpreted as due to the onset of a spin density wave with the same period as the CDW in presence of a prominent electron-phonon interaction originating from the Rashba spin-orbit coupling.\\
While it is difficult to make definitive statements about the generality of our observations, we anticipate that by tuning the stoichiometric properties and doping level of topological insulators - thus changing the position of the Dirac point and possible nesting conditions - the condition for CDW order may be changed or shifted to a different periodicity. It is thus expected that from further experiments and validation one may be able to evolve phase diagrams for Dirac CDWs as a function of stoichiometry, doping and Fermi level position. Taken together the results promise also to shed light on previous experimental and theoretical investigations of related systems and how CDW order affects lattice dynamics and stability.

\section*{Acknowledgement}
This research was funded in whole, or in part, by the Austrian Science Fund (FWF) [J3479-N20, P29641-N36 \& P34704-N]. For the purpose of open access, the author has applied a CC BY public copyright licence to any Author Accepted Manuscript version arising from this submission. We thank Marco Bianchi for his advice and help in terms of the sample preparation and additional characterisation of the samples. We would also like to thank Philip Hofmann (Aarhus University), Evgueni Chulkov (DIPC, San Sebastian, Spain) and D. Campi \& M. Bernasconi (Unimib, Milano, Italy) for many helpful discussions.

\section*{Additional information}
The online version contains supplementary material.

\bibliography{literature}

\clearpage

\begin{widetext}
\begin{center}
\textbf{\large Supplementary Information:\\ Observation of Dirac Charge Density Waves in Bi$_2$Te$_2$Se\\}
\end{center}
\end{widetext}

\setcounter{section}{0}
\renewcommand{\thesection}{S\arabic{section}}

\renewcommand{\thepage}{\arabic{apppage}}
\pagenumbering{arabic}
\setcounter{table}{0}
\setcounter{figure}{0}
\setcounter{figure}{0}
\makeatletter
\renewcommand{\thesection}{S\arabic{section}}
\renewcommand{\theequation}{S\arabic{equation}}
\renewcommand{\thefigure}{S\arabic{figure}}

\section{Experimental Details}
\label{sec:experimental}
The experimental data of this work were obtained at the HAS apparatus in Graz, where a nearly monochromatic beam of $^4$He is scattered off the sample surface in a fixed source-sample-detector geometry with an angle of $91.5^{\circ}$ (for further details about the apparatus see Ref. \cite{Tamtogl2010}). The scattered intensity of a He beam in the range of $10-20$ meV is then monitored as a function of the incident angle $\vartheta_i$ with respect to the surface normal, which can be modified by rotating the sample in the main chamber (base pressure $p \leq 2\cdot 10^{-10}~\mbox{mbar}$). The momentum transfer parallel to the surface $\Delta K$, upon elastic scattering, is given by 
$\Delta K = \lvert \mathbf{k}_i \rvert  \left( \sin \vartheta_f  - \sin \vartheta_i  \right)$, with $\mathbf{k}_i$, the incident wavevector, and $\vartheta_i$ and $\vartheta_f$ the incident and final angle, respectively.\\
Details about the sample growth procedure can be found in Mi \textit{et al.} \cite{Mi2013}. \SA\ exhibits a rhombohedral crystal structure, which is in accordance with other bismuth chalcogenides built of quintuple layers (QL, see Figure 1 in main text) which are bound to each other via weak forces of van der Waals character \cite{Mi2013}. The conventional hexagonal unit cell consists of 3 QLs with each QL layer being terminated by a Te layer (Figure 1(a) in main text). The weak bonding between the QLs gives access to the (111) cleavage plane with the described Te termination and a surface lattice constant of $a=4.31~\mbox{\AA}$ (Figure 1(b) in main text).\\
The \SA\ sample was fixed on a sample holder using thermally and electrically conductive epoxy and cleaved \textit{in situ}, in a load-lock chamber \cite{Tamtogl2016a}. The sample temperature can be varied by cooling via a thermal connection to a liquid nitrogen reservoir and heating based on a button heater. After cleavage, the cleanliness and purity of the sample can be further checked using low energy electron diffraction (LEED) and Auger electron spectroscopy (AES). 
\begin{figure}[htbp]
     \centering
     \centering
	\includegraphics[width=0.45\textwidth]{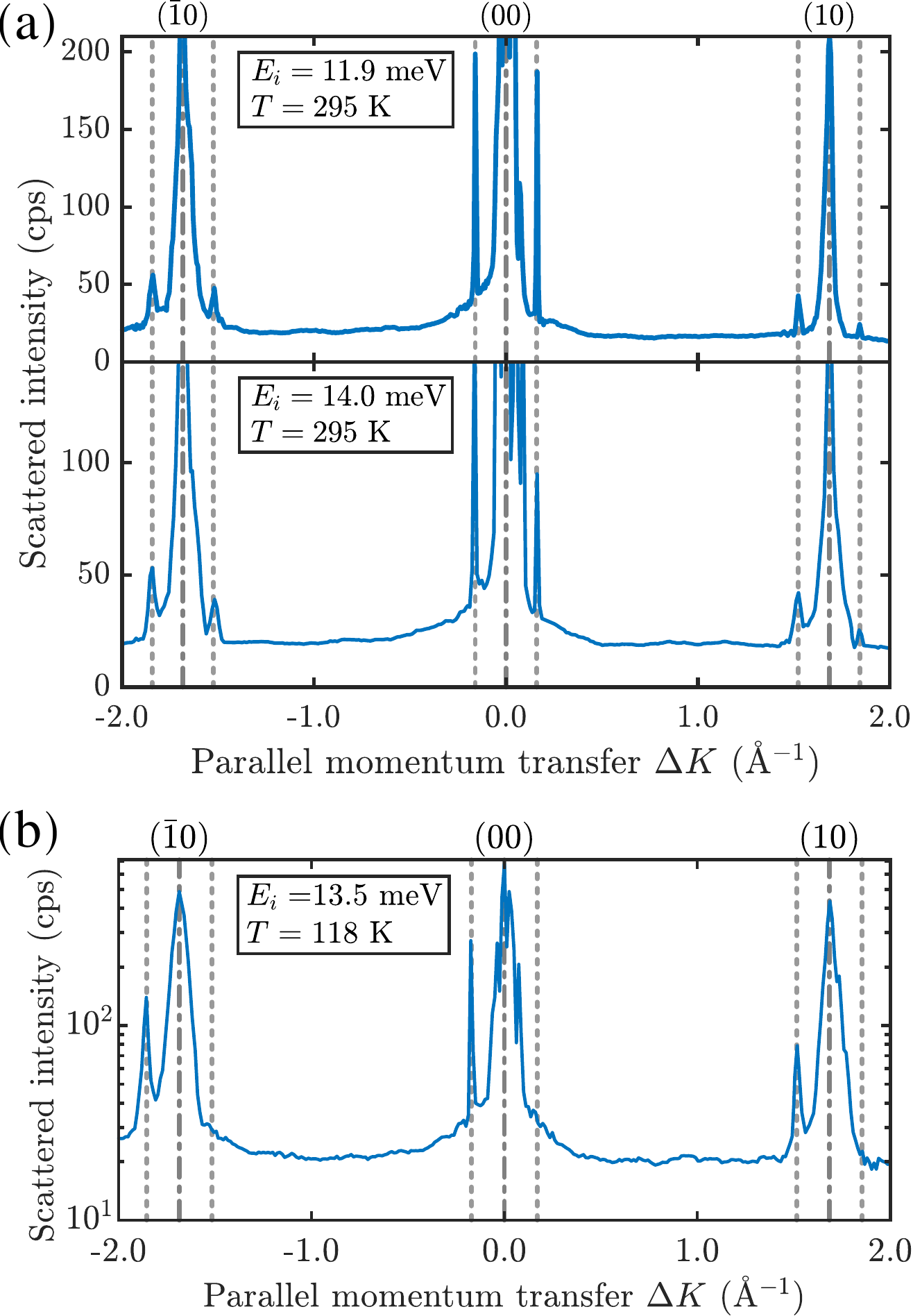}
	\caption{(a) Additional full diffraction scans taken at room temperature and with varying incident beam energy $E_i$ further illustrate the spin-allowed CDW periodicity of \SA(111) along \GM . The corresponding satellite peaks (grey dashed vertical lines) appear next to the specular and first order diffraction peaks (grey dash-dotted vertical lines). (b) An additional full diffraction scan (logarithmic scale) with the sample cooled down illustrates that the asymmetry between $+ \mathbf{g}_{\mathrm{CDW}}$ and $- \mathbf{g}_{\mathrm{CDW}}$ as described in the main text is also evident for the satellite peaks next to the first order diffraction peaks.}
     \label{fig:FigS2}
\end{figure}

\section{CDW satellite peaks and anomalous Debye attenuation}
\label{sec:SI_CDW}
Both the position of the peaks assigned as CDW and the described temperature dependence cannot be attributed to a lack of cleanliness or to a superstructure formed by adsorbates at the surface. While we cannot exclude that the structure very close to the specular may be caused by twinning / different domains, which can occur on layered crystals such as the binary TIs\cite{Kevy2021}, such an effect cannot cause the appearance of the $\pm \mathbf{g}_{\mathrm{CDW}}$ satellite peaks next to both the specular and the diffraction peaks.
\begin{figure}[htbp]
	\centering
	\includegraphics[width=0.45\textwidth]{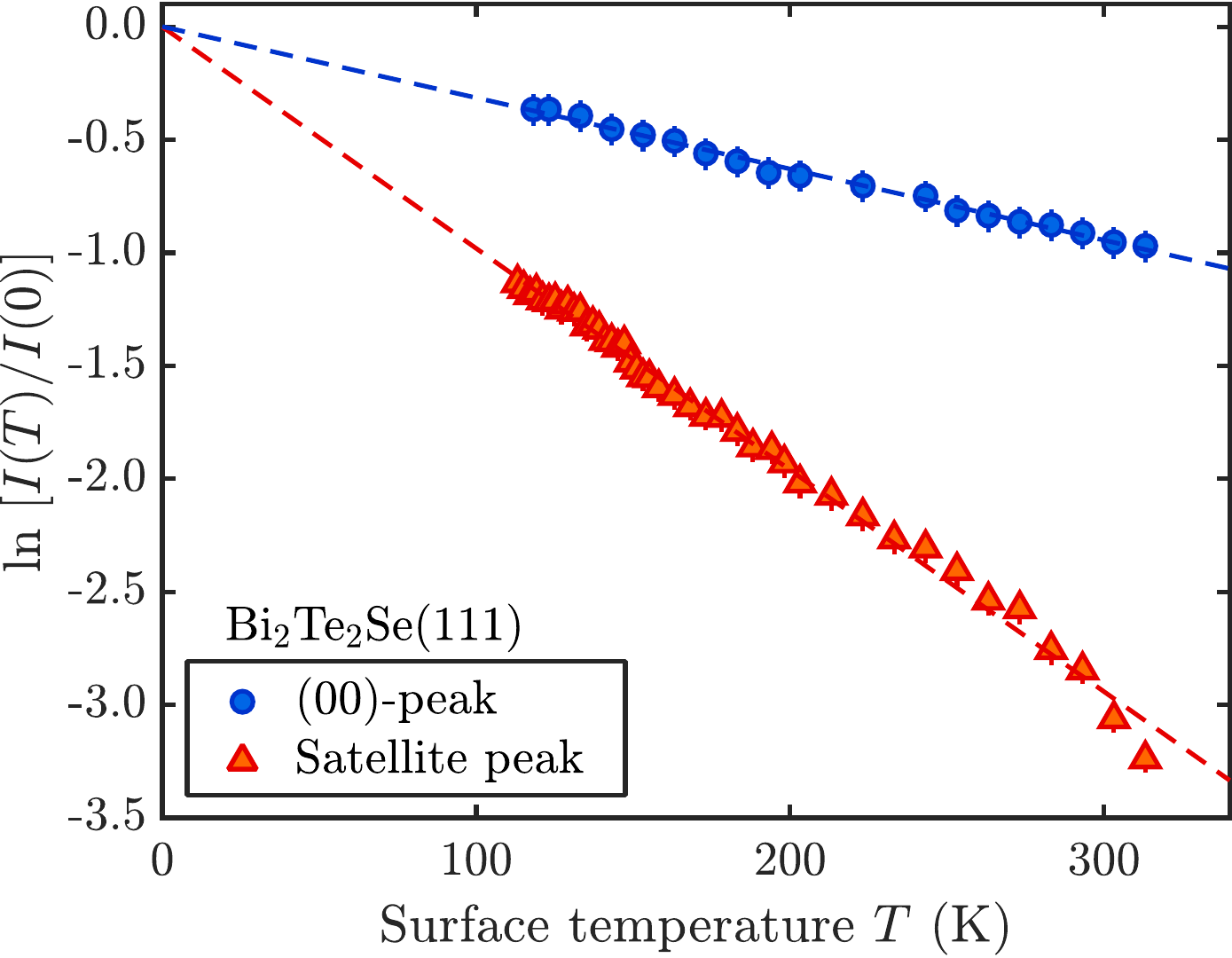}
	\caption{Debye-Waller plot of the temperature dependence for both the specular as well as the satellite peak at the left hand side of the specular peak.}
	\label{fig:S1}
\end{figure}

Therefor, \autoref{fig:FigS2} shows a number of additional full diffraction scans with varying incident beam energies. The full scan in \autoref{fig:FigS2}(b), illustrates that the asymmetry between $+ \mathbf{g}_{\mathrm{CDW}}$ and $- \mathbf{g}_{\mathrm{CDW}}$ for the cooled sample, as described in the main text, is also evident for the satellite peaks next to the first order diffraction peaks. Small intensity variations occurring between the right- and left-hand side of the specular may be present due to misalignment effects, however, \autoref{fig:FigS2}(b) shows that such artefacts cannot explain the asymmetry between $+ \mathbf{g}_{\mathrm{CDW}}$ and $- \mathbf{g}_{\mathrm{CDW}}$. It should further be noted, that the CDW peaks are very sensitive to the azimuthal orientation of the crystal, with the spin-allowed CDW periodicity of \SA(111) occurring only along \GM , which clearly speaks against specular scattering. The temperature dependence of the peak intensities with the typical critical exponent is further strong evidence for the satellite peaks originating from a CDW.\\
Finally, the semi-lograthmic plot (\autoref{fig:S1}) of the obtained temperature dependence for the diffraction intensities of the specular reflection (blue) and the satellite peak (red), clearly indicates the anomalous attenuation of the CDW satellite peak compared to the Debye attenuation of the specular peak. The data plotted in \autoref{fig:S1} is further used for the derivation of the order parameter in Figure 4(b) in the main text.

\end{document}